\documentclass[a4paper,fleqn]{cas-dc}

\usepackage[authoryear]{natbib}

\def\tsc#1{\csdef{#1}{\textsc{\lowercase{#1}}\xspace}}
\tsc{WGM}
\tsc{QE}
\tsc{EP}
\tsc{PMS}
\tsc{BEC}
\tsc{DE}

\begin{document}
\let\WriteBookmarks\relax
\def\floatpagepagefraction{1}
\def\textpagefraction{.001}

\shorttitle{Prostate Cancer Segmentation with PCa-RadHop}

\shortauthors{Vasileios Magoulianitis et~al.}

\title [mode = title]{PCa-RadHop: A Transparent and Lightweight Feed-forward Method for Clinically Significant Prostate Cancer Segmentation}                      




%
\author[1]{Vasileios Magoulianitis}[
                        orcid=0000-0001-7511-2910]

\cormark[1]


\ead{magoulia@usc.edu}



\affiliation[1]{organization={Electrical and Computer Engineering Department, University of Southern California (USC)},
    addressline={3740 McClintock Ave.}, 
    city={Los Angeles},
    postcode={90089}, 
    state={CA},
    country={USA}}

\affiliation[2]{organization={Department of Urology, Keck School of Medicine, University of Southern California (USC)},
    addressline={1975 Zonal Ave.}, 
    city={Los Angeles},
    postcode={90033}, 
    state={CA},
    country={USA}}

\affiliation[3]{organization={Department of Radiology, Keck School of Medicine, University of Southern California (USC)},
    addressline={1975 Zonal Ave.}, 
    city={Los Angeles},
    postcode={90033}, 
    state={CA},
    country={USA}}

\author[1]{Jiaxin Yang}[]
\author[1]{Yijing Yang}[]
\author[1]{Jintang Xue}[]
\author[2]{Masatomo Kaneko}[orcid=0000-0002-1205-807X]
\author[2]{Giovanni Cacciamani}[orcid=0000-0002-8892-5539]
\author[1]{Andre Abreu}[orcid=0000-0002-9167-2587]
\author[3,2]{Vinay Duddalwar}[orcid=0000-0002-4808-5715]
\author[1]{C.-C. Jay Kuo}[]
\author[2]{Inderbir S. Gill}[orcid=0000-0002-5113-7846]
\author[1]{Chrysostomos Nikias}[]

\cortext[cor1]{The corresponding author is with the Electrical Engineering Department of University of Southern California (USC), Los Angeles, USA.}



\begin{abstract}
Prostate Cancer is one of the most frequently occurring cancers in men, with a low survival rate if not early diagnosed. PI-RADS reading has a high false positive rate, thus increasing the diagnostic incurred costs and patient discomfort. Deep learning (DL) models achieve a high segmentation performance, although require a large model size and complexity. Also, DL models lack of feature interpretability and are perceived as ``black-boxes" in the medical field. PCa-RadHop pipeline is proposed in this work, aiming to provide a more transparent feature extraction process using a linear model. It adopts the recently introduced Green Learning (GL) paradigm, which offers a small model size and low complexity. PCa-RadHop consists of two stages: Stage-1 extracts data-driven radiomics features from the bi-parametric Magnetic Resonance Imaging (bp-MRI) input and predicts an initial heatmap. To reduce the false positive rate, a subsequent stage-2 is introduced to refine the predictions by including more contextual information and radiomics features from each already detected Region of Interest (ROI). Experiments on the largest publicly available dataset, PI-CAI, show a competitive performance standing of the proposed method among other deep DL models, achieving an area under the curve (AUC) of 0.807 among a cohort of 1,000 patients. Moreover, PCa-RadHop maintains orders of magnitude smaller model size and complexity.
\end{abstract}



\begin{keywords}
Biparametric MRI \sep Prostate cancer segmentation \sep  Data-driven radiomics \sep Feed-forward model \sep Interpretable pipeline
\end{keywords}

\maketitle

\section{Introduction}\label{sec:introduction}
Prostate Cancer (PCa) is widely known as one of the most frequently occurring cancer in diagnosis in men. Reportedly, in 2020, it accounts for more than 1.4 million diagnosed cases (7.3\% of new cancer
cases), while 375,304 people deceased from PCa \citep{miller2019cancer}. Among other cancers, it was the fifth leading cause of death in the same year. An important fact according to the American Cancer Society \citep{siegel2023cancer} is that the five year survival rate when early diagnosed reaches almost 100\%, whereas in metastasis stage that number plummets to 31\%. As a consequence, early diagnosis of clinically significant cancer (csPCa) plays a vital role in patients life. 



Currently, abnormally elevated serum prostate-specific antigen (PSA) and positive digital rectal examination (DRE) are two indicative factors for a patient to undergo biopsy for detecting csPCa. Yet, both methods tend to overdiagnosis, incurring higher health costs and patient discomfort (such as erectile dysfunction and urinary incontinence). As of 2019, to increase the sensitivity and specificity of detecting csPCa, European Association of Urology (EAU) guidelines state that multiparametric magnetic resonance imaging (mpMRI) is recommended as the initial diagnostic test prior to biopsy. This procedure has radiologist to examine MRI sequences and assign with a PIRADS score \citep{weinreb2016pi}. Along this procedure, a high false positive rate and discordance among radiologists in detecting csPCa has been reported among different studies \citep{seo2017pi,kasel2016assessment}. The accurate segmentation of lesions is very significant, as it affects the accuracy of MRI-targeted biopsies and also touches upon focal therapy \citep{o2021future, piert2018accuracy}, in order to increase therapy’s effect on tumor region, while minimizing the radiation dose on noncancerous tissue. 

In recent years and since the advent of Deep Neural Networks (DNNs), the research in Computer-aided Diagnosis (CAD) tools has been largely increased. DNNs have enabled tasks and applications, such as fully automated prostate gland and lesion segmentation, as well as PIRADS classification. In turn, the automation of these tasks gives rise to a fully automated prostate cancer detection pipeline, without any human intervention. DL-based works use a backbone architecture for segmentation tasks \citep{jaeger2020retina,vnet} (e.g. U-Net, V-Net) and further built up new modules on top of it. Attention-mechanisms \citep{duran2022prostattention, duran2020prostate, yuan2022z} have been also employed to boost the regional-features from the two main zones of prostate, since they have different visual characteristic according to studies \citep{ginsburg2017radiomic}. Some works target suspicious ROIs detection (PIRADS-based) \citep{sanford2020deep}, while other methods try to detect csPCa \citep{abraham2018computer, yoo2019prostate} or stratify even further the associated Gleason score of each lesion \citep{cao2019joint, abraham2018computer_medical}.

Despite their high performance at various computer vision tasks, DNNs are often criticized as a “black-box” methodology for high-stake decisions. Deep features are hard to interpret how they are derived and in fact have no physical meaning to experts. Future CAD tools should be trustworthy from physicians, so they can be employed in a real clinical setting. 
Green Learning (GL) paradigm \citep{kuo2022green, chen2020pixelhop++} provides an alternative pattern recognition and feature learning framework, offering a more transparent multi-scale feature extraction using the Successive Subspace Learning (SSL) approach. GL employs simple signal processing techniques and a linear model to create a rich feature space, by decomposing the input image into a spatial-spectral representation \citep{chen2020pixelhop++, rouhsedaghat2021successive}. Besides, GL offers a very lightweight model solution comparing to the DNNs, thus reducing the overall complexity of the task. This work decouples from the main body of DL-based solutions for csPCa detection and builts upon the GL framework.

The overall pipeline consists of two stages: Stage-1 yields an initial heatmap with a probability of each voxel harboring csPCa,  trying to capture parts of the lesion area in a patch-wise manner and using GL for feature extraction. Across the rich spatial-spectral representation, csPCa is more discriminant on certain dimensions that are selected using the Discriminant Feature Test (DFT) before training the classifier. Additionally, stage-2 grows a larger area about candidate detections from Stage-1, since csPCa lesions need a larger context to be disentangled from false positives \citep{yu2020deep, yu2020false}. Stage-2 key aim is to reduce the csPCa probability of false positive (FP) detections and increase that of true positives (TP). It uses both probability-based and visual-based features, that is, hand-crafted radiomics, combined with GL-based radiomics features from stage-1.

In this work, we propose novel prostate cancer detection method with a transparent feature extraction process and very small model size, comparing to other state-of-the-art works. The main contributions can be summarized in four folds:
\begin{enumerate}
\item To the best of our knowledge, PCa-RadHop is the first work applies the Green learning paradigm and SSL methodology for a cancer detection task. 
\item A novel two stages pipeline method is proposed, where each module is intuitive and transparent. 
\item An linear unsupervised, data-driven radiomics-like feature extraction method is proposed, named RadHop.
\item Benchmark experiments and ablation studies are conducted on the currently largest publicly available PI-CAI dataset.
\end{enumerate}

The rest of the paper is organized as follows. Related work is reviewed in Section~\ref{sec:related_work}. The proposed method is presented in Section~\ref{sec:methods}. The experimental analysis is discussed in Section~\ref{sec:results} and conclusions are drawn in Section~\ref{sec:conclusion}.

\begin{figure*}[th]
\begin{center}
\includegraphics[width=1.0\linewidth]{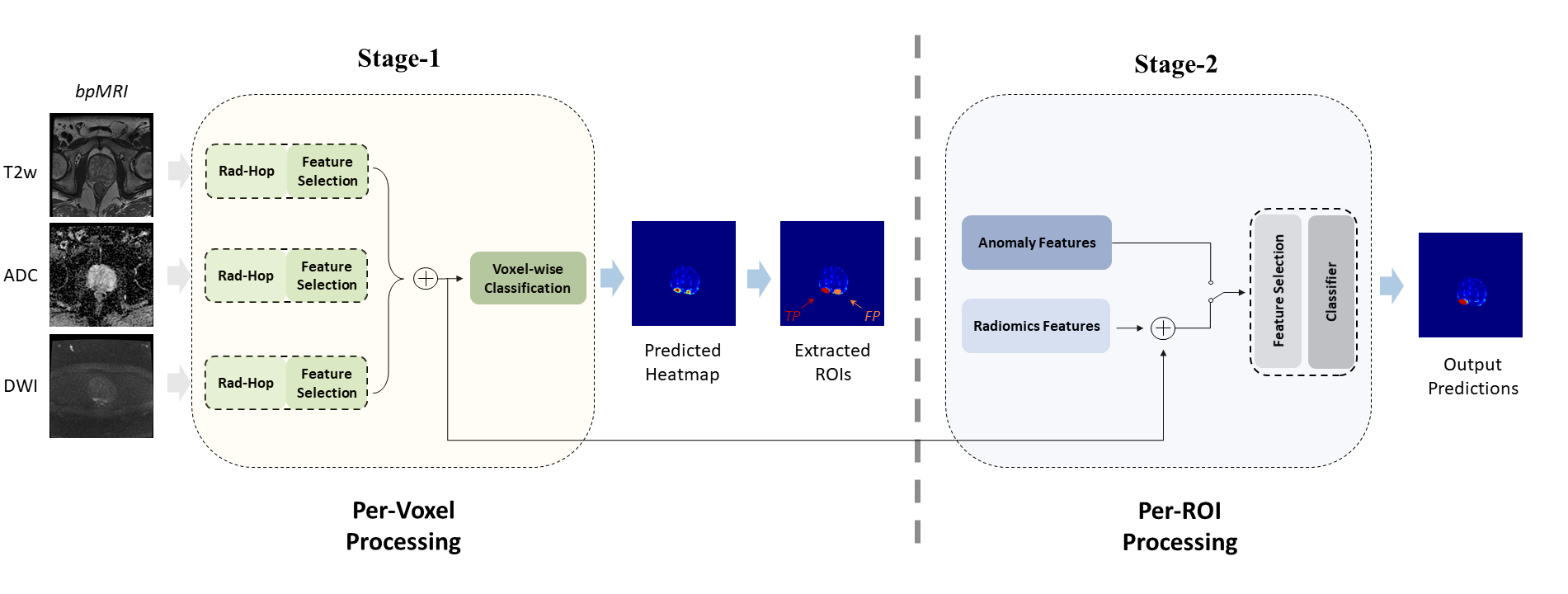}
\end{center}
\vspace{-5mm}
\caption{The overall PCa-RadHop pipeline is illustrated. In stage-1 (yellow box), the per-voxel data-driven feature extraction and selection process take place independently for each of the three sequences. Then, selected features are concatenated before classifier's input, where a probability heatmap is obtained. In stage-2 (blue box), detected ROIs are further processed for discerning FPs, by combining anomaly features that provide more contextual probability, as well as hand-crafted and data-driven radiomics.}
\label{fig:overall_pipeline}
\end{figure*}

\section{Related Work}\label{sec:related_work}

\subsection{PCa Detection \& Lesion Segmentation}

Since lesion detection can be viewed also as a semantic segmentation task, most of the works use popular segmentation networks to identify and segment suspicious ROIs. U-Net \citep{unet, jaeger2020retina} is a popular choice in literature for various semantic segmentation task and thereby can be used for identifying suspicious areas on the prostate gland  \citep{siddique2021u, wong2021fully, huang2021application, hambarde2020prostate, isensee2021nnu, sanyal2020automated, khosravi2021deep}. \cite{huang2021application}, employ a U-Net to extract a weight map and propose a novel fusion mechanism for T2-w and ADC sequences based on Gaussian and Laplacian pyramid decomposition. \cite{wong2021fully} propose an ensemble of several U-Net based models with different architectures. A recent work \citep{wen2023ipca} proposes a detection method for incidental prostate cancer, which is the early stage before clinically significant cancer.

Another work \citep{hambarde2020prostate} uses T2-w input only and a radiomics-based supervised U-Net for prostate gland and lesion segmentation. It demonstrates that radiomics-based pipeline improves segmentation accuracy over U-Net. Moreover, nnU-Net model was proposed \citep{ isensee2021nnu} meant to offer a more generic solution for medical image segmentation tasks. This method is self-configured with respect to preprocessing, network architecture and training parameters. A comparative study for detecting csPCa \citep{schelb2019classification} compares U-Net predictions using bpMRI input with PI-RADS scoring, reaching to a conclusion that automated AI-based predictions achieve similar performance with PI-RADS reading from radiologists.

Attention mechanisms are widely adopted in computer vision tasks and are proved to increase the performance. ProstAttention-Net is proposed by \cite{duran2022prostattention} utilizing U-Net as backbone architecture and has two branches, one for gland and the other for lesion segmentation. The latter one uses the predicted prostate mask as a prior through an attention mechanism for detecting and further grading lesions. A similar approach is also proposed in \citep{duran2020prostate} for steering network's attention on the Peripheral Zone of the prostate. On the same track \citep{yu2020deep} uses ResNet-50 as backbone and a feature pyramid network (FPN) to efficiently combine global and local features.



\cite{cao2019joint} propose the FocalNet that is trained using the focal loss (FL) in order to mitigate the class imbalance between negative and positive areas, and mutual finding loss (MFL) on T2-w and ADC to extract the most discriminant cross-modality features. FL is also adopted in \citep{cao2019prostate} for lesion segmentation, using as post-processing a selective dense conditional random field to refine the initial lesion segmentation. \cite{zhang2021cross} propose a cross-modal self-attention distillation network for learning complementary non-linear information across different modalities. 

\cite{shao2021patient} propose an end-to-end method for efficiently combining the patient-level ground truth with the slice-level predictions during training. Earlier, \cite{wang2018automated} had introduced a way to train a CNN for csPCa in a weakly supervised way, using only image-level annotations. 

\subsection{Lesion Classification} \label{subsec:les_clasf_related}
A few other works decouple from jointly segmenting lesions and aim at classifying the identified ROIs in terms of the histopathological report. \cite{de2020deep} introduce a 2D U-Net trained to predict the Gleason grade group, using ordinal encoding and also incorporating prostate zonal mask as prior information. Prostate gland crops are used in \citep{yang2017co}, to extract a cancer response map per slice using a fully convolutional network. Inconsistency loss controls the prediction discrepancies between T2-w and ADC sequences. Another interesting work \citep{yu2020false} proposes a multi-scale false positive reduction module. A pre-DL popular approach for classifying an ROI and trying to quantify visual features from cancer molecular patterns that reflect on MRI is radiomics \citep{gillies2010biology, lambin2012radiomics}. A number of works \citep{cuocolo2019clinically, algohary2018radiomic, yao2020radiomics, toivonen2019radiomics} have proved that radiomics can help discerning csPCa from suspiciously looking ROIs. \cite{cameron2015maps} use hand-crafted radiomics to detect certain attributes on the prostate, by maintaining interpretability for radiologists.


\subsection{Successive subspace learning methodology}

As already seen, most of the problem approaches base their pipelines on a DCNN model trained using back-propagation. A new framework, Green Learning (GL), has been recently introduced from \cite{kuo2022green}, aiming to provide a more transparent feed-forward feature extraction process, carrying a small number of parameters and complexity. In particular, the successive subspace learning (SSL) methodology was proposed in a sequence of papers \citep{ kuo2018data,kuo2019interpretable}. Drawing some parallel lines with deep learning, it creates a multi-scale feature extraction scheme by adding layers interleaved with pooling operations, to trade spatial for spectral features \citep{chen2020pixelhop++}. Instead of convolutional filters trained with backpropagation, principal component analysis (PCA) is used to learn the local subspace across different layers, where each feature has also a larger receptive field along deeper stages. Each layer is called Hop, in which features are learned in an unsupervised data-driven way.  In practice, the input signal is decomposed via the subspace approximation via adjusted  (Saab) transform into a rich spatial-spectral representation \citep{kuo2019interpretable}. 

Within the medical field, the first two works using GL have been proposed from \cite{liu2021voxelhop, liu2021segmentation} on cardiac segmentation and ALS disease classification. The proposed pipeline uses the channel-wise Saab transform for voxel-wise feature extraction and classification. To the best of our knowledge, this is the first work that applies the GL paradigm and SSL framework to cancer detection.

\begin{figure*}[t]
\begin{center}
\includegraphics[width=1.0\linewidth]{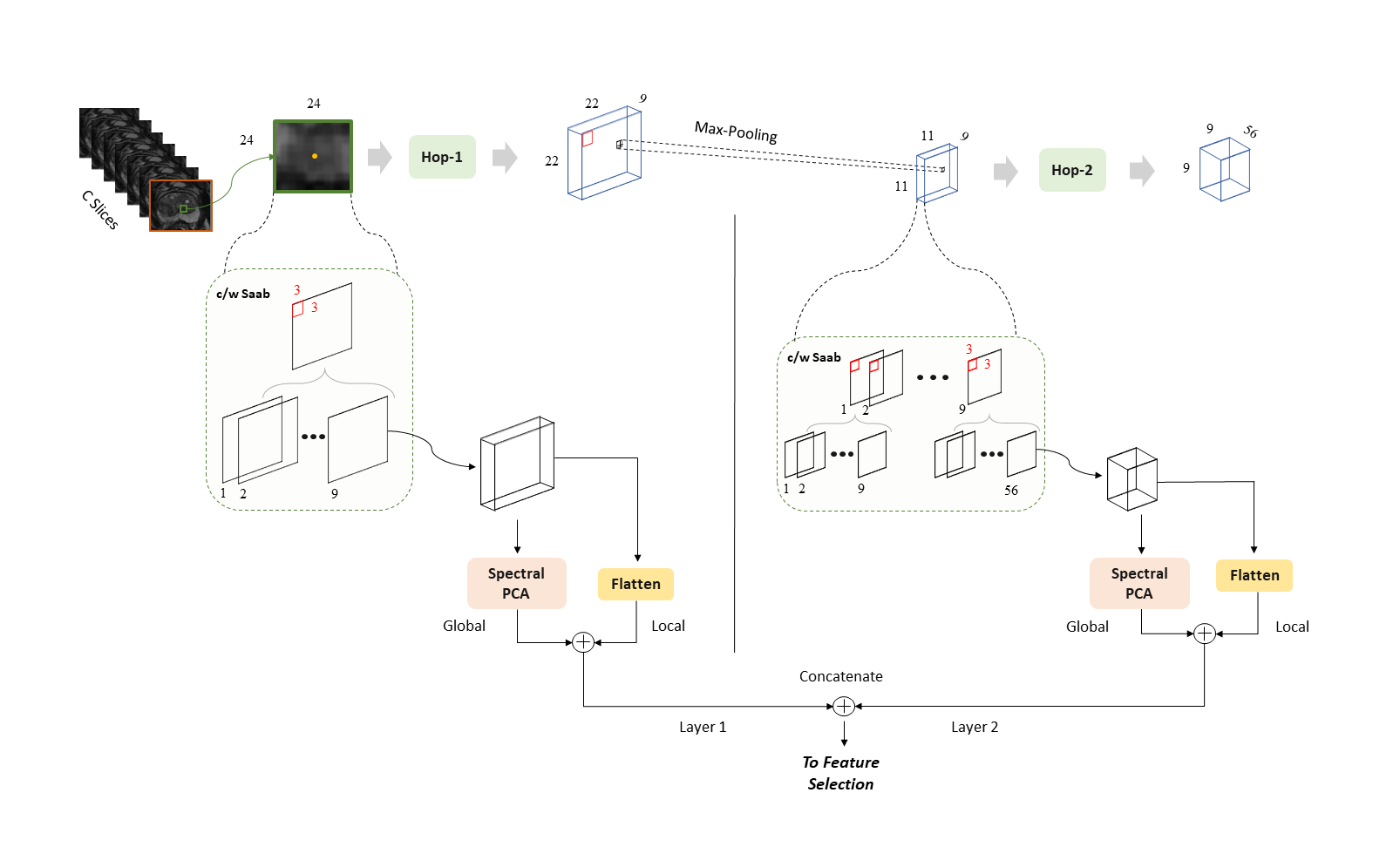}
\end{center}
\vspace{-5mm}
\caption{Unsupervised feature extraction with RadHop pipeline. A patch centered at a certain location is fed in RadHop. Two concatenated Hop units bring multi-scale properties in the final feature representation. Local features have spatial information, while global ones refer to the overall feature map for each spectral dimension. The output feature is the concatenation from the local and global features of the two layers.}
\vspace{-3mm}
\label{fig:RadHop}
\end{figure*}

\section{Methodology}\label{sec:methods}
The proposed method receives as input a bp-MRI sequence (T2-w, ADC and DWI high b-value) and as a pre-preprocessing step all sequences are sampled on T2-w dimensions. It consists of two stages: (1) In the first stage (Section \ref{subsec:stage_1}), 2D voxel-wise predictions are made to deduce the csPCa probability of each voxel. The classification unit operates on a fixed patch size --surrounding a voxel-- used for feature extraction following the SSL methodology and classification. Then, the anomaly score map is derived in Section \ref{subsec:anomap} that provides initial indication of suspiciously looking areas with high probability of harboring csPCa. One question arises, what is the optimal patch size needed to discern csPCa. Prior research \citep{sanyal2020automated, khosravi2021deep} suggests a patch size between $21$ to $31$.

Our proposed PCa-RadHop detection framework has been devised to classify initially large parts of the lesion, adopting a patch size of $24\times24$ in stage-$1$. Then, we locally aggregate initial predictions to acquire $8\times8$ blocks and construct the anomaly map. In stage-$2$, operating on $8\times8$ units we grow a larger area about each candidate region, considering an area of $40\times40$ for classification, since recent research suggest for false positive reduction, a larger area is required about each potential lesion \citep{yu2020false}. An overview of the proposed pipeline is illustrated in Figure~\ref{fig:overall_pipeline}.

\subsection{Anomaly Detections (Stage-1)} \label{subsec:stage_1}

This stage yields initial predictions per voxel, targeting a relatively smaller patch size than the average lesion area. Certainly, stage-$1$ can work as a standalone module and directly generate a voxel-wise heatmap of csPCa. Two patch sizes are experimented in stage-$1$ (see Section \ref{sec:results}): (1) $24\times24$ for calculating the anomaly map and feed stage-2 input, and (2) as an independent module for predicting csPCa operating on $32\times32$ patch size. 
 
\subsubsection{Data-driven Radiomics -- Feature Extraction} 

This is the core feature extraction module of our detection framework, based on SSL. Off-the-shelf GL-based choices for feature extraction vary depending on the task. We opt for E-PixelHop \citep{yang2021pixelhop} method to borrow our ideas about the feature extraction architecture and tailor it to MRI input specifications. The feature extraction module itself within PCa-RadHop pipeline is named Rad-Hop, as it extracts radiomic-like features from MRI input in a data driven way. It consists of two consecutive steps: 1) neighborhood construction (per-slice), and 2) representation learning through channel-wise Saab transform~\citep{chen2020pixelhop++}.

There are two kind of features within Rad-Hop, the spatial (or local) and spectral (or global) ones. The former represent certain local parts of the ROI, whereas the spectral ones have a global view of the input feature map of each layer. This spatial-spectral decomposition is motivated by the assumption that csPCa is expressed more on certain spectral components and hence through feature selection the most discriminant subspaces are retained for classification. Figure~\ref{fig:RadHop} demonstrates the architecture and components of the proposed RadHop feature extraction module that has two layers of Hop units. For MRI signal input, adding more layers adds more complexity without significant performance improvement.

The input tensor in RadHop is of dimension $H_i\times W_i\times C_i \times K_i$, where $H_i$, $W_i$ and $C_i$ represent the resolution in the 3D space, and $K_i$ represents the dimension of the feature vector for each voxel extracted from the $(i-1)$-th RadHop unit ($K_i$ projected subspaces). Since our csPCa detection pipeline is based on 2D predictions, $C=1$, and square patches are extracted from single slices of size $S_i$, that is, $H_i=W_i=S_i$. At the first layer of RadHop  (Hop-$1$), $K_0=1$ and thus the input tensor is of dimension $S_1\times S_1\times 1 \times 1 $ for each MRI sequence. We first gather the local neighborhood in the 2D space centered at each voxel on the prostate gland area. The neighborhood size within Hop-i is defined as $F_{i}\times F_{i}\times 1$ (independent for each $K_i$ dimension) and denotes the local filter size. That is, each voxel in the neighborhood has corresponding feature vector of dimension $K_i$, which results in a tensor of size $F_{i}\times F_{i}\times K_i$ in Hop-i output. Channel-wise Saab transform is performed on each of the $K_i$ subspaces independently, to derive the spectral kernels through PCA for each subspace of the previous layer. 

\subsection{Subspace Approximation in Green Learning}

Within SSL framework, PCA extracts the orthogonal subspaces of local neighborhoods and in turn each subspace image is further decomposed from the next Hop unit. On each layer this can be expressed as the following affine transform:
\begin{equation}\label{eq:affine_transform}
y_m = \mathbf{a}_m^T\cdot \mathbf{x} + b_m, \hspace{4mm}m=0,1,\cdots, M-1,
\end{equation}

where $\mathbf{x}$ is the input tensor of size $\mathbf{x}\epsilon \mathbb{R}^{{Fi}\times {Fi}}$, representing the local neighborhood, $\mathbf{a}_m$ is the $m$-th anchor vector (i.e. eigenvectors of PCA), and $M$ is their total number. The channel-wise Saab transform \citep{chen2020pixelhop++} decomposes the input subspaces into the direct sum of two subspaces, the DC and AC, as expressed in Eq.~\ref{eq:SDC_SAC}.

\begin{equation}\label{eq:SDC_SAC}
\hspace{20mm} S = S_{DC} \oplus S_{AC} .
\end{equation}

\begin{figure*}[t]
\begin{center}
\includegraphics[width=1.0\linewidth]{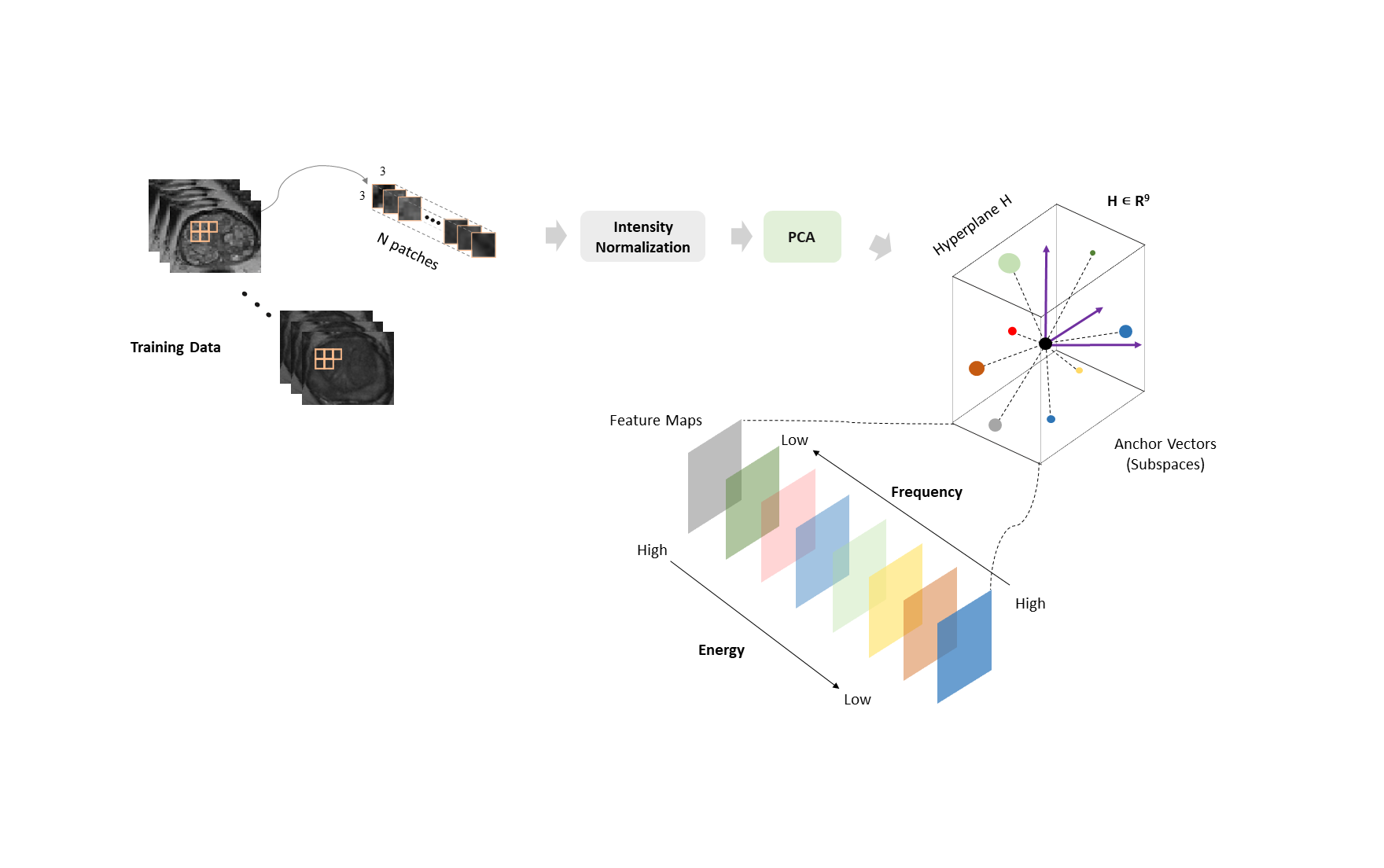}
\end{center}
\vspace{-20mm}
\caption{Illustration of the feature decomposition into spectral components using the Saab transform in SSL. PCA is used to identify the orthogonal subspaces. Different subspaces have different energy (shown with different color size). This is the core module for feature extraction in Hop unit employed from RadHop. Malignant areas reflect differently on certain subspaces and hence RadHop feature representation can help the classifier to detect them.}
\label{fig:subsapce}
\end{figure*}

$S_{DC}$ is defined as the DC anchor vector, given from $\mathbf{a}_0=\frac{1}{\sqrt{N}}\left ( 1, 1, \cdots , 1 \right )^T$, while the $S_{AC}$ is spanned by the $\mathbf{a}_m$ anchor vectors. The two subspaces are orthogonal to each other, where the input signal $\mathbf{x}$ is projected onto $\mathbf{a}_0$ to get the DC component $\mathbf{x}_{DC}$. Then the AC component is extracted by subtracting the DC component from the input signal, i.e. $\mathbf{x}_{AC}=\mathbf{x}-\mathbf{x}_{DC}$. Then, AC anchor vectors are learnt by conducting PCA on the AC component. The first $K$ principal components with sufficient energy are kept as the AC anchor vectors. Hence, features can be extrated by projecting $\mathbf{x}$ onto the above learnt anchor vectors (i.e. eigenspace) based on Eq.~\ref{eq:affine_transform}. The bias term is derived according to the following work~\citep{kuo2019interpretable}. The subspace decomposition out of local filters from the training data is illustrated in Fig.~\ref{fig:subsapce}.

In RadHop, we use two Hop units (see Fig. \ref{fig:RadHop}). Both Hop-1 and 2 use a filter size of $F_{i} = 3$ to learn the anchor vectors. All spatial projections on a certain subspace are assembled on a feature map $f_{s}^{(i)}$ of Hop-$i$ and subspace $s=0,1,\cdots, K_{i}$. Without using padding on the input tensors in each layer, the output feature map in Hop-1 is of size $(S_0-1)\times (S_0-1)\times K_1$. Prior to Hop-2, a max pooling by $2$ is applied to reduce the spatial feature map dimension. Hop-2 spatial dimension output is $(S_1/2-1)\times (S_1/2-1)\times K_2$. $K_i$ corresponds to the number of spectral components (projected eigenspaces) and equals to the input dimensions of PCA ($K_1 = F_{1} \times F_{1} \times K_{0}$). 

In Hop-2, according to c/w Saab transform \citep{chen2020pixelhop++, kuo2019interpretable} each feature map is further projected onto several anchor spaces independently from the rest feature maps (i.e. spectral dimensions). $K_2$ equals to the aggregated number of all subspaces in Hop-2 which equals $F_2 \times F_2 \times K_1$. Therefore, the maximum number of feature maps in Hop-1 and 2 is $9$ and $81$ respectively (given $F_1 = F_2 = 3$). All feature maps are orthogonal to each other. 

Certainly, not all spectral components are equally informative for the input signal distribution. Those with very few energy as determined from the eigenvalue (always positive) magnitude are discarded. Especially for signals like MRI, this can lead to a much smaller model size, without much loss of information.


C/w Saab transform in each Hop within RadHop decomposes the input signal into a different multi-scale representation where each feature in Hop-i, $f_{x,y}^{(i)}$ has dimension $1 \times 1 \times K_i$ corresponding to a specific spatial position (local features).

Yet, desiring to extract also features that have a global view of the local patch, we further use PCA on each of the $K_i$ feature maps of each Hop layer, based on the $S_i \times S_i \times 1$ spatial responses. In doing so, for each feature map $f^{(i)}$ of size $S_i \times S_i \times K_i$, we extract one global feature using PCA of feature dimension $S_i \times S_i$. After PCA, the feature map is decoupled from the spatial domain and we obtain the spectral one. Global features $G_{s}^{(i)} \in \mathbb{R}^{S_i \times S_i}$ can be defined as:
 
\begin{equation}\label{eq:global_feats}
\hspace{18mm}   G_{s}^{(i)} = PCA(f_{s}^{(i)}) \hspace{2mm} \forall s, \forall i.
\end{equation}
 
The final feature output from RadHop results from the concatenation of flattened feature maps from both Hop units (local features), as well as their corresponding global features. In RadHop, DC is also included along with the AC filters, as it is important for csPCa classification.

\subsubsection{Feature Selection} 

RadHop provides a rich spatial-spectral representation of the input patch. However, even after filtering out feature maps with very low energy, the feature dimensions is still very high for training a classifier. Furthermore, some dimensions may carry very little discriminant power for classifying csPCa. RadHop operates in a completely unsupervised way, where the feature space is very rich, having a physical meaning in signal processing terms --by encoding different texture patterns of the MRI signal-- but a high portion of the feature space is not useful for csPCa classification. The goal of the feature selection module is to extract the most discriminant feature subspace from RadHop, thus providing a less noisy input to the classifier.

Within GL framework, a novel Discriminant Feature Test (DFT) has been proposed by \cite{yang2022supervised} to quantify the discriminant power of features. A brief high level description is given below: For a given feature $a$ under test, the minimum and maximum of projected values $f(a) = a^{T}x$ is calculated, denoted by $f_{min}$ and $f_{max}$, respectively. The partition interval
$[fmin, fmax]$ is split into $B$ bins in a uniform way. For measuring the discriminant power and class separation power of each features the bin boundaries are used as candidate thresholds. Each threshold, $t_{b}$, $b = 1$,...,$B - 1$ partitions
interval $[fmin, fmax]$ into two sub-intervals. Using as candidate thresholds all the bin boundaries, the splitting quality can be evaluated using the weighted entropy:

\begin{equation}
    L_{a, t_b} = \frac{N_+}{N_+ + N_-} L_{a,t_{b+}} + \frac{N_-}{N_+ + N_-} L_{a,t_{b-}}
\end{equation}

where $N_+ =| F_{a,t_{b}+} |$ and $N_{-} =| F_{a,t_{b}-} |$ is the samples number in
the left and right sub-intervals, respectively. As a cost function the entropy value is chosen to measure the purity of the partition:

\begin{equation}
 \hspace{20mm}   L = - \sum_{c=1}^{C} p_{c}log(p_{c})
\end{equation}

Across all $L_{a, t_b}$ values the minimum cost is chosen to represent the discriminant power of the given feature. Running the DFT across all features from RadHop, we then sort them in ascending order according to their minimum entropy values. A natural number of features to keep with the lowest entropy value is at the elbow point that usually occur from the distribution of the sorted RadHop features. That is, from each MRI sequence $1,000$ features are kept. The three features from the overall MRI input are concatenated to finalize the fused feature  $X$ as input to the classifier (see Fig.~\ref{fig:overall_pipeline}).

\begin{equation}\label{eq:F_s}
\hspace{18mm}   X = \left [X_{T2}, X_{ADC}, X_{DWI} \right ] 
\end{equation}

\subsubsection{Feature Classification}

The last part of stage-$1$ is to train a classifier to perform voxel-wise predictions --similar to this pipeline \citep{giannini2015fully}--, given the extracted feature $X$ from RadHop for each voxel, after the late stage concatenation of the three sequence features of the bpMRI. We opt for the Xtreme Gradient Boosting (XGB) classifier \citep{xgb} because it is more powerful from a Random Forest classifier, since the decision trees are not independent, but each new tree is built upon prior tree errors through the loss function. Also, XGB offers a small model size solution that does not compromise our objective for an energy efficient pipeline for csPCa. 

\subsubsection{Hard Negative Mining}


A commonly known issue in PCa localization and detection, in general, is the class imbalance problem, since there are numerous negative areas on the prostate gland, but just a handful of ROIs with csPCa. Given that in most studies usually the patient cohort is also imbalanced, as most of the patients have no malignant findings or if any they are indolent (non csPCa). As such, one can realize the problem occurring with the class imbalance problem. Another issue, most of the negative samples are very easy to be classified because most of the prostate gland areas are not suspicious looking \citep{yeung2022unified}. Therefore, by randomly sampling patches for training, gives the impression to the classifier that negative samples are far away from the positive ones in feature space. Nevertheless, there are still many prostate regions closer to the positive class that can cause false positives, thereby affecting detector's specificity.  

Towards this end and trying to mitigate this issue, we perform a two-step training of the classifier. In the first step, having no other choice, we train the XGB using random patch sampling. We densely sample within positive ROIs patches with a 25\% overlap from each annotated slice (especially when patch is of size $24\times24$ where ground truth lesion is partially captured). Negative regions are randomly picked across prostate and their number is selected accordingly to the desired class ratio. The optimal ratio, negative to positive patches, for training the XGB was empirically found to give the best performance at roughly 4:1. 

After step-$1$ training, XGB is applied almost densely (with some small stride) all over the prostate glands in the training cohort and a ``soft" label is obtained, which is the classifiers out-of-bag probability for a given sample. One can realize that the negative class distribution in the patient cohort is very skewed. Most of the negative samples will be picked will be far from the positive class distribution and that would make the classifier prone to false positives. In a hard negative mining manner, we train XGB again in a second step, having now the prior for each patch where it stands within feature space (``easy" or ``hard" based on the soft label). To balance out the training in step-$2$, we split the soft label distribution into $10$ bins. The number of training samples is determined from the overall number of positive areas. Given that it is desired to maintain a certain class ratio, the negative samples population can be decided. Finally, we apply an exponential rule to decide how many samples $N_{i}$ will be sampled out of each ``hardness" bin shown in Eq.~\ref{eq:negative_mining}. 

\begin{equation}\label{eq:negative_mining}
\hspace{7mm} N_{i} = N e^{-\lambda(i-1)}, \hspace{10mm} i = 0,...,9
\end{equation}
 
 where $N$ the total number of negative samples and $i$ the bin index. Following the exponential distribution we include more easier samples and as the difficulty increase in term of soft labels, the number of samples decays exponentially. Hyper-parameter $\lambda$  was set at $0.4$. The XGB after the second training round is used from the detector to provide per voxel predictions and calculate the csPCa heatmap.

\begin{table}[t]
\centering
\caption{Architecture of the proposed RadHop Unit }
\label{tab:encoder_archi}
\begin{tabular}{cccc}
\toprule
 &  Input Resolution & Filter Size &  Stride \\
\midrule
Hop 1      & $(24\times24)\times1$  & $(3\times3)\times1$ &  $(1\times1)\times1$ \\
Max-pool 1   & $(22\times22)\times1$ &  $(2\times2)\times1$ &  $(2\times2)\times1$ \\
Hop 2      & $(11\times11)\times1$ &  $(3\times3)\times1$ &  $(1\times1)\times1$ \\
\bottomrule
\end{tabular}
\end{table}


\subsection{Anomaly Map Calculation}\label{subsec:anomap}

\begin{figure}[t]
\begin{center}
\includegraphics[width=1.0\linewidth]{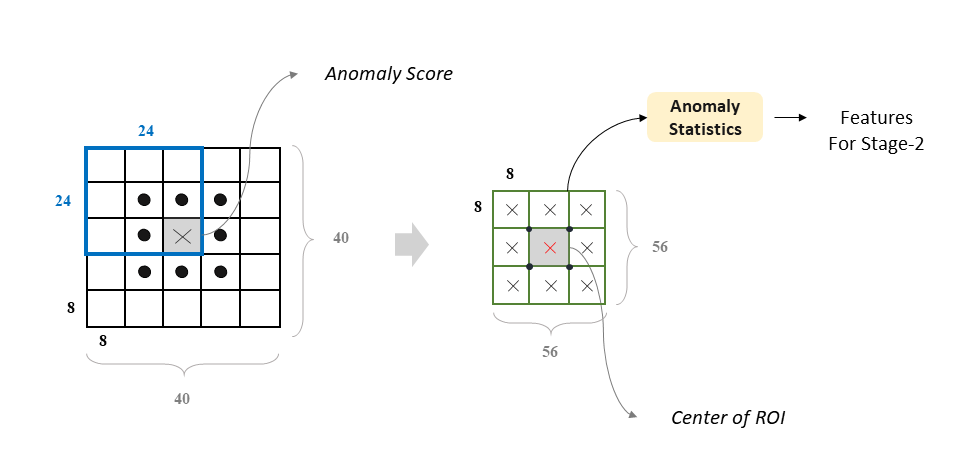}
\end{center}
\vspace{-5mm}
\caption{Anomaly map ($\Lambda$) calculation from stage-1 predictions. Nine neighboring voxel predictions are averaged using $24\times24$ patches with stride 8 on stage-1 output. Then, in stage-2 the anomaly map is used to extract statistical features from the neighboring anomaly score distribution.}
\label{fig:anom_map}
\end{figure}

Having obtained initial predictions for each voxel in stage-1, we aim to grow a larger area around suspiciously looking regions. Voxel-level probability predictions are more sensitive to errors. Therefore, it is desired to make units that are more robust, by aggregating neighboring predictions. To this end, average pooling is applied about each voxel using a $3\times3$ area. For patch size of $24\times24$, if we consider averaging $9$ surrounding patches with stride equals $8$, then the receptive field for the center voxel area of $8\times8\times1$ is $40\times40\times1$ (see Fig. \ref{fig:anom_map}). 

The center $8\times8$ unit is an overlap area for all $9$ neighboring patches. Hence, moving the process now into stage-2 our processing units are no longer the voxels, but $8\times8$ unit blocks conveying an anomaly score that results from averaging neighboring voxel-level predictions and the span on the voxel space is large enough to help distinguish false positives from true positives. We name the array of anomaly scores ($\Lambda$), which is a subsampled mapping of the original input resolution.  

\subsection{Anomaly Candidates Classification (Stage-2)}\label{subsec:decoder}

As mentioned before, stage-1 can predict a heatmap on the prostate gland with probabilities for areas harboring csPCa. Hence, one can use it as a separate module for cancer detection and lesion segmentation, choosing appropriately the processing patch size for feature extraction. In the experimental section we show results for a larger and smaller patch size. For stage-2 inputs, we opt for a patch size of $24\times24$ in order to capture smaller parts of the average lesion area.

The main object of stage-2 is to grow a larger area around high anomaly scores from the previously calculated anomaly map. Larger patch size may include more parts of the lesion but given lesion's irregular shape can also include more noise. Using smaller patches we can predict on parts from the lesion and in stage-2 to combine predictions from the neighborhood to classify true positive areas over false positive ones with high anomaly score.

\subsubsection{Candidate Areas For Stage-2} \label{subsec:candidates}

Having calculated the anomaly map $A$, where each anomaly score corresponds to an $8\times8$ area, one needs to decide what candidate regions are sent to stage-2. That is, the only available information for deciding that so is the anomaly scores. It is also important to ensure that not many true positive (TP) areas are missed from stage-2 processing and also to quickly remove some anomaly scores that are very unlikely to be true csPCa lesions. 

To this end, we binarize $A$ at the anomaly threshold (T) where the true positive rate (TPR) is very high (i.e. close to 1) and the FPR is minimized. Having binarized $A$, one can use some simple rules for deciding which candidates areas continue to stage-2. Single anomaly scores, even more robust than a voxel, they can still correspond to some small false positive regions that is always likely to exist somewhere. Therefore, instead of one anomaly score, after binarization we use a $2\times2$ element to detect candidates on the anomaly map. This corresponds to a $16\times16$ area in the original MRI space. In this way, many small regions, unlikely of being TPs can be filtered out, thus reducing noisy information in stage-2. The optimal anomaly score found to be $T_{opt} = 0.38$ when minimizing the FPR, while constraining the TPR to be over a certain point ($r = 0.95$) (see Eq.\ref{eq:F1_optimal}).

\begin{equation}\label{eq:F1_optimal}
\hspace{5mm} T_{opt} = {\operatorname{argmin}} \underset{T} (FPR) \hspace{5mm} s.t. \hspace{5mm} TPR \ge r
\end{equation}

\subsubsection{Feature Extraction - Anomaly-based}

The last step and main object of stage-2 is to classify candidate regions with high anomaly score over the binarization threshold and retain a $2\times2$ structure. Usually the number of false positive (FP) candidates is more than the TPs. Hard sample mining technique aims at reducing the FP predictions and is important for stage-2 to provide a cleaner input. 

One way to build features for training another classifier in stage-2 and grow a larger area surrounding the candidate anomaly units --which is the main motivation to adding a second stage-- is to include the $9$ neighboring anomaly units (non-overlapping) in a feature vector, plus their mean and standard deviation. Given that each anomaly units is of size $8\times8$ the whole area we consider is of size $56\times56$ in the original MRI input (see Fig.~\ref{fig:anom_map}). Finally, a classifier can be trained on TPs and FPs as a second classification step.

\subsubsection{Handcrafted Radiomics}

Having identified the suspicious candidates from the anomaly map, we can still fall back on the original input MRI sequence and extract additional visual-based features. Radiomics library \citep{lambin2012radiomics} provides a good solution for quantifying local characteristics of an ROI. They are usually employed after having determined an ROI and used to classify it with respect to the biopsy report.

As long as stage-1 has provided the initial predictions, ROIs can be drawn either using directly the generated heatmap from stage-1 or following the anomaly map candidates \ref{subsec:candidates}. For classification using radiomics we consider features from all categories within radiomics library, except for the Shape-based 3D (i.e $n=104$ for each sequence). Moreover, we enhance the radiomics features with RadHop features from stage-1, using a standard window size of $24 \times 24$ from the center of mass of the ROI. DFT module is also employed here to select the most discriminant set of radiomic features, in case some extracted features are more noisy. 

\begin{table}[t]
\centering
\caption{Radiomics Features Categories}
\label{tab:complexity}
\begin{tabular}{cc}
    \toprule
      Feature Type & Num. of Features \\
      \midrule
      First Order Statistics & 19 \\
      Shape-based (3D) & 16 \\
      Shape-based (2D) & 10 \\
      Gray Level Co-occurence & 24 \\
      Gray Level Run Length & 16 \\
      Gray Level Size Zone & 16 \\
      Neighbouring Gray Tone Difference & 5 \\
      Gray Level Dependence & 14 \\
\bottomrule
\end{tabular}
\end{table}

\section{Experimental Setup}\label{sec:experiments_setup}

\subsection{Database and pre-processing}

To demonstrate the effectiveness of the proposed csPCa method using Green Learning, we conduct experiments based on the currently largest publicly available dataset from the PI-CAI challenge \citep{saha2023artificial}, which consists bpMRI data of a cohort of $1,500$ patients. A subset of  $328$ cases are common with the previously largest  dataset provided at the ProstateX challenge \citep{armato2018prostatex}. Besides, an online hidden validation set of $100$ patients is offered. For the testing phase, all methods were evaluated once on a cohort of $1000$ patients. 

Training data comprise MRI scans across multiple clinics, such as the Radboud University Medical Center (RUMC), University Medical Centet Groningen (UMCG) and Ziekenhuis Groep Twente (ZGT). Scans from those medical centers are included both in training and testing. MRI scans from the Norwegian University of Science and Technology (NTNU) is also included in the testing cohort, as an unseen institute data to training. The vendors of MRI scanners are Siemens Healthineers and Philips Medical Systems. The training data are divided into $1075$ samples of benign or indolent PCa and $425$ clinically significant cancer cases. Prior RadHop, all MRI sequences are resampled onto T2-w dimensions using resampling.   


\subsection{Evaluation Metrics}

For measuring the performance of the proposed pipeline, as well as benchmarking with other existing works, same metrics as in the PI-CAI challenege are adopted \citep{saha2023artificial}. That is, there are two metrics to evaluate the methods performance in detecting csPCa. For lesion-level detection the average precision (AP), where each segmented lesion has a floating point probability for being clinically significant. For patient-level performance, Area Under Receiver Operating Characteristic (AUROC) is used, predicting for each patient a probability of having csPCa. For the PI-CAI challenge the average between the two metrics was used to rank different methods. For the per-lesion performance, to consider a hit with the TP lesion, the prediction must have an Intersection over Union (IoU) at least $0.1$. 

\begin{equation}\label{eq:Precision}
  \hspace{20mm} P = \frac{T_p}{T_p + F_p} 
\end{equation}

\begin{equation}\label{eq:Recall}
  \hspace{20mm} R = \frac{T_p}{T_p + F_n} 
\end{equation}

\begin{equation}\label{eq:Precision_Recall}
  \hspace{13mm}  AP = \sum\limits_{n} (R_n - R_{n-1})P_n
\end{equation}

 
\begin{equation}
   \hspace{11mm} Score = \frac{(AP + AUROC)}{2}.
    \label{eq:picai_score}
\end{equation}

\section{Experimental Results}\label{sec:results}
\subsection{Results on PI-CAI Dataset}
For benchmarking with other methods we use the results from the PI-CAI challenge on the testing cohort of patients. For ablation study and intermediate results between stage 1 and 2, a local validation cohort is randomly chosen, consisting of $300$ patients. For a fair comparison, we report results with baseline methods from PI-CAI challenge trained solely based on the PI-CAI publicly available cohort, without using external private datasets. The performance in both metrics and comparisons are available in Table~\ref{tab:Overall_Results}.  

\begin{figure*}[t]
\begin{center}
\includegraphics[width=0.9\linewidth]{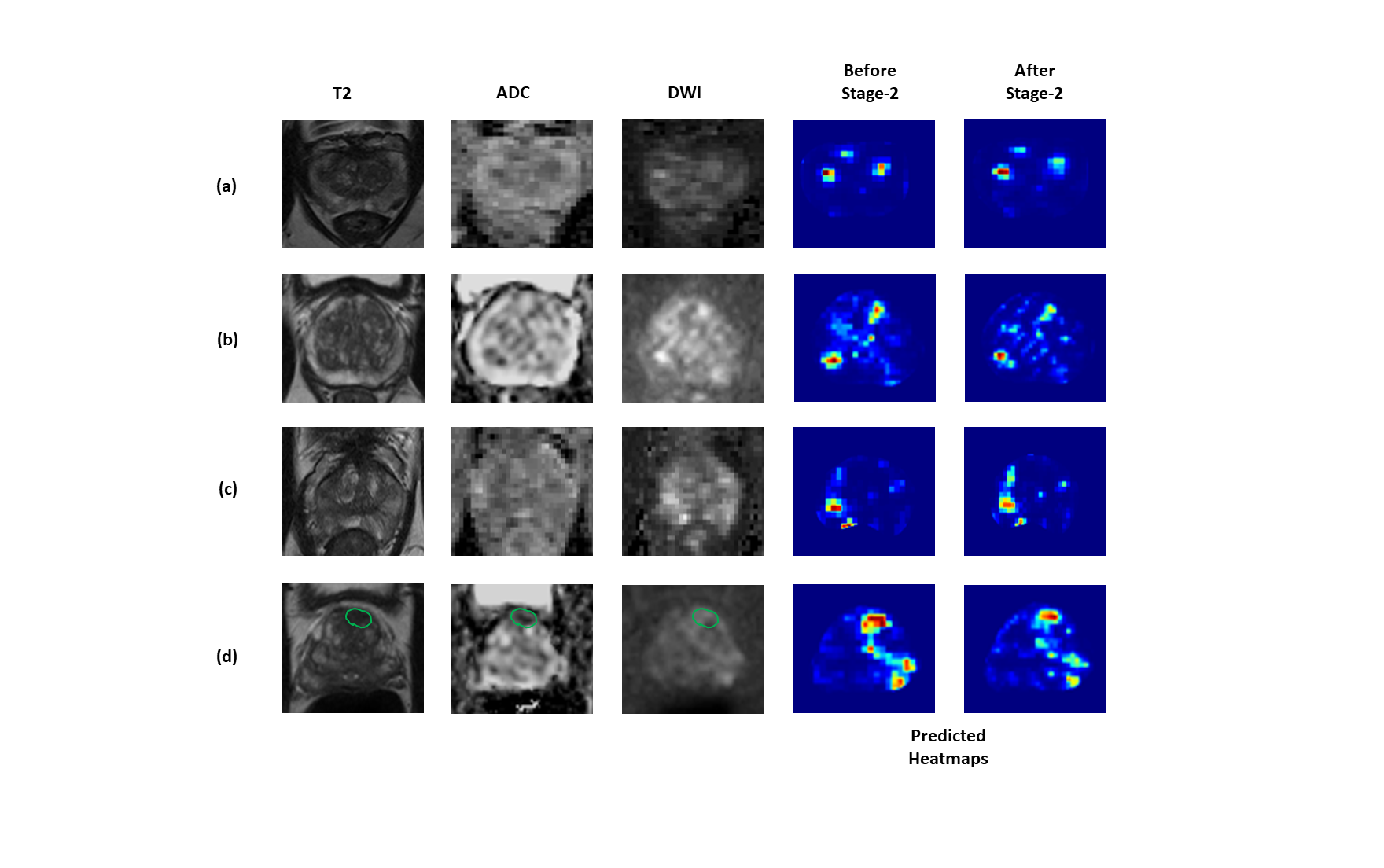}
\end{center}
\vspace{-15mm}
\caption{Visualizations from csPCa-RadHop intermediate predictions in stage-1 and the false positive reduction in stage-2. In the three first rows (a)-(c) three negative cases are shown with some false positive areas and how their probability is decreased after stage-2. In the last row (d) a positive case is displayed. The true positive ROI (green) is retained after stage-1, while the probability of the other false positive ROIs is reduced.}
\label{fig:qualitative_results}
\end{figure*}

\begin{figure*}[t]
\begin{center}
\includegraphics[width=0.8\linewidth]{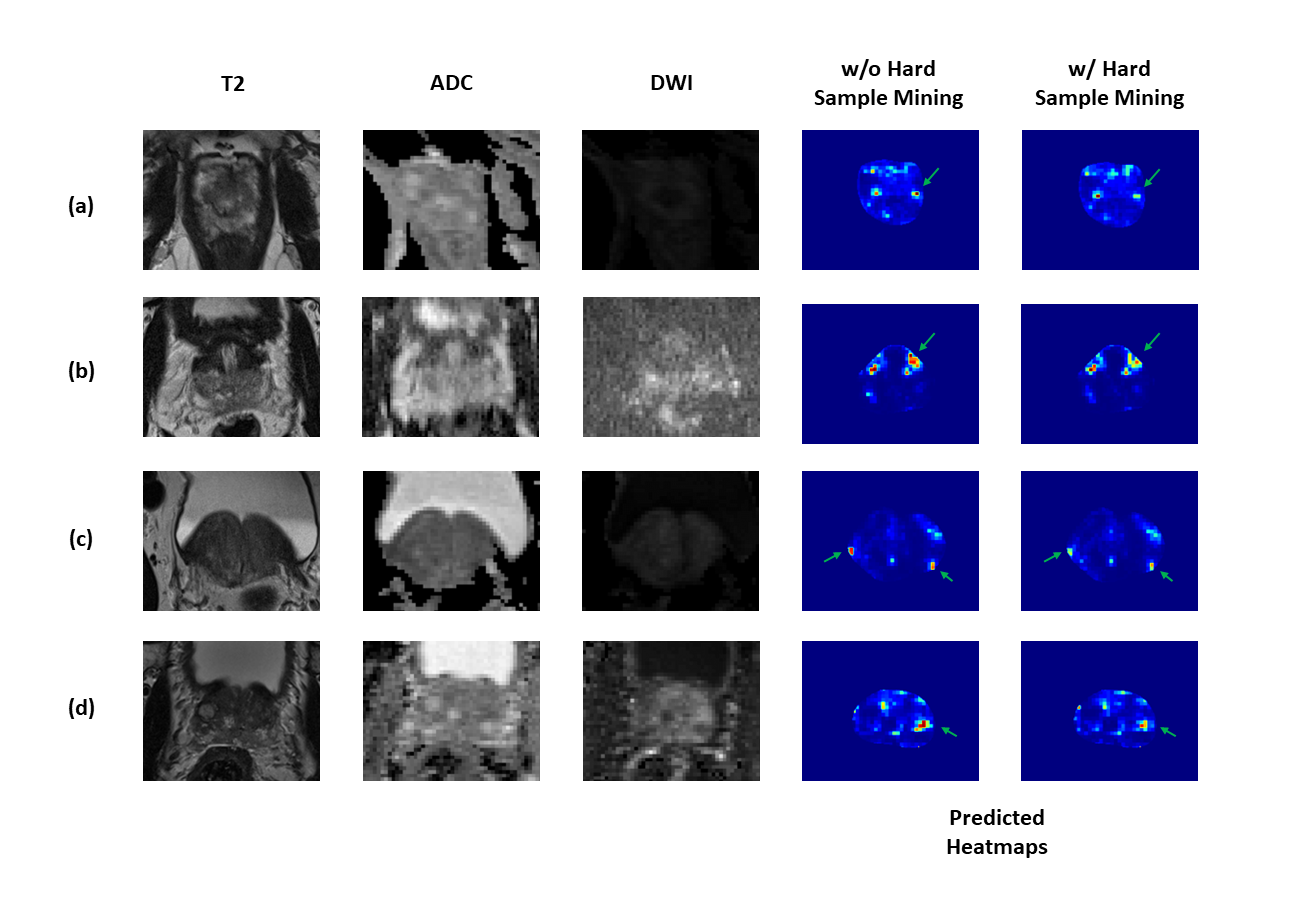}
\end{center}
\vspace{-5mm}
\caption{Demonstration of the hard sample mining technique effectiveness on four negative examples from different patients with areas prone to false positives. The predicted heatmaps from two differently trained classifiers is shown for each case. Without hard sample mining implies that the classifier is trained on randomly chosen patches across patients. The green arrows point to specific FP regions where hard sample mining increases the robustness of the classifier and reduces the probability of certain FPs.}
\label{fig:qualitative_results_mining}
\end{figure*}

As we can see RadHop including only stage-1 predictions using patch size of $32\times32$ (i.e. no refinement by stage-2), it has a competitive standing among other methods using DCNN architectures, such as U-Net or Swin-Transformer. The evaluation in such a large patient cohort minimizes the bias and approaches better the actual performance, should our algorithm used in an actual clinical setting. Both the patient-level prediction metric (i.e. AUROC) and lesion-level precision (i.e. AP), surpass the performance of nnDetection and Swin-Transformer models. In patient-level, RadHop has a better performance than every other method under comparison, expect U-Net that has a slightly better performance. In lesion-level metric, RadHop achieves a higher precision than other methods, but there is still some gap with the U-Net related ones. This observation after error analysis motivated us to introduce a stage-2 step, to refine the heatmap predictions from stage-1. In Fig.~\ref{fig:qualitative_results}, we provide heatmap visualizations about stage-2 effectiveness in reducing the FP probability from stage-1 predictions. By reducing the FP probabilities --wherever possible-- using the classifier in stage-2, both the case-level (AUROC) and lesion-level (AP) scores increase. Turns out that adding more contextual probability information information through the anomaly map and combining with radiomics features (both data-driven and hand-crafted) helps to decrease the probability (specificity) of certain FPs, without compromising the TP rate (sensitivity). In Figure~\ref{fig:qualitative_results_mining}, it is also shown the classifier robustness in FP, when trained with the hard sample mining technique rather than random sampling.

\begin{table}[t]
\centering
\caption{Performance benchmarking with selected DL-based models based on 1,000 testing patients from PICAI challenge.}
\label{tab:Overall_Results}
\begin{tabular}{c|c|c|c}
\toprule
      Method & Score   & AUROC & AP \\ 
\midrule
SwinTransformer & 0.561 &  0.729 & 0.393 \\ 
nnDetection     & 0.586 &  0.785 & 0.386  \\ 
nnU-Net \citep{isensee2021nnu}         & 0.626 &  0.803 & 0.450 \\ 
U-Net \citep{jaeger2020retina}          & 0.635 &  0.814 & 0.456  \\ 
PCa-RadHop (Stage-$1$ only) & 0.607 & 0.807  & 0.407  \\ 
\bottomrule
\end{tabular}
\end{table}

\subsection{Ablation Study}

\begin{figure}[t]
\begin{center}
\includegraphics[width=1.0\linewidth]{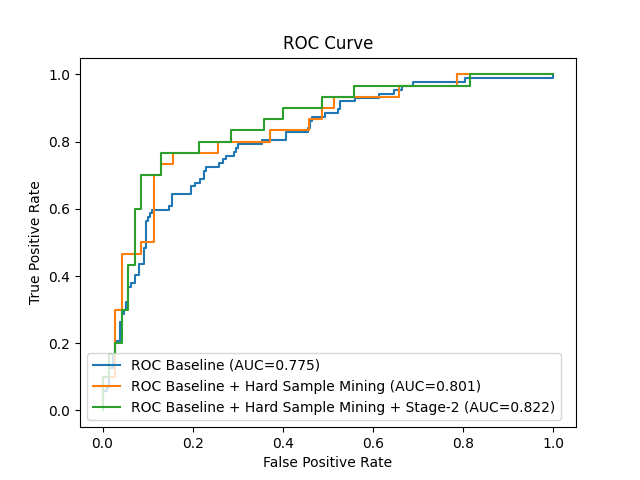}
\end{center}
\caption{ROC curves and AUC comparison of three different pipelines, showing the effectiveness of the hard sample mining technique and stage-2 in reducing the false positive rate.}
\label{fig:AUROC_Ablation}
\end{figure}

Given the limited access to the hidden testing cohort of PI-CAI challenge, we perform a performance comparison to demonstrate the effectiveness of stage-2 and compare stage-1 in two different patch sizes. Table \ref{tab:Ablation}, shows some experiments carried out, trying to improve stage-1 output. At first, one can see the effectiveness of using smaller patch size, since it captures more easily parts of the lesion, as opposed to try to identify the lesion as a whole. For doing the latter, one needs a larger patch size that can induce a more noisy RadHop feature representation, thereby affecting the precision performance, especially for medium-smaller lesion sizes. Stage-2 effectiveness is also evident in Table \ref{tab:Ablation}. In particular, AP (lesion-level metric) is improved from $0.356$ to $0.374$, while the AUROC (patient-level metric) increases from $0.801$ to $0.822$ (see Figure~\ref{fig:AUROC_Ablation}).This is due to the fact that it is possible to reduce the probability of some false positives in stage-2, by re-classifying ROIs detected in stage-1, using the anomaly features, as well as the hand-crafted and data-driven (RadHop) radiomics features. 

\begin{table}[t]
\centering
\caption{Results on a local validation dataset ($300$ patients) and performance comparison after applying stage-2 processing.}
\label{tab:Ablation}
\begin{tabular}{c|c|c|c}
\toprule
      Configuration & Score   & AUROC & AP \\ 
\midrule
Stage-1 ($32 \times 32$) & 0.558 &  0.768 & 0.348   \\ 
Stage-1 ($24 \times 24$) & 0.579 &  0.801 & 0.356   \\
Stage-1 ($24 \times 24$) + Stage-2 & 0.599 &  0.822 & 0.374  \\ 
\bottomrule
\end{tabular}
\end{table}

\subsection{Model Size and Complexity Comparison}

An important aspect of models comparison, besides performance, is the model size and the Floating Point Operations (FLOPS). These two factors are consequential when algorithms are implemented on certain platforms, especially with limited computation resources. Table \ref{tab:complexity} shows the comparisons with nnU-Met and U-Net architectures (popular choices among several methods in literature for PCa detection). One can realize the tiny model size of RadHop when compared to DL-based architecture ($\times1445$ and $\times1681$) more parameters for U-Net and nnU-Net respectively). In addition, the computational complexity of RadHop is fairly low. Specifically, nnU-Net requires $\times128$ operations than RadHop for inference, while U-Net requires $\times56$ FLOPS. These findings highlight the tremendous deployment advantages of our method when compared with DL models. Also, this stresses one of the Green Learning framework key benefits which is to offer lighweight and environment-friendly solutions. Additionally, RadHop has a transparent feed-forward pipeline, where each module functionality can be explained to physicians.   

\begin{table}[t]
\centering
\caption{Model size and complexity comparison.}
\label{tab:complexity}
\begin{tabular}{ccc}
\toprule
      Model & Model Size (M) & FLOPS (B) \\ 
\midrule
nnU-Net &  $185$ $(\times 1681)$ & $346$ $(\times 128)$ \\ 
U-Net & $159$ $(\times 1445)$ &  $152$ $(\times 56)$ \\  
{\bf PCa-RadHop (Ours)} & {\bf0.11} $(\times 1)$ &  {\bf 2.7} $(\times 1)$ \\ 
\bottomrule
\end{tabular}
\end{table}

\section{Conclusion}\label{sec:conclusion}

A fully automated two stages pipeline for csPCa detection, PCa-RadHop, is proposed in this work, aiming to offer a more transparent framework for feature extraction and analysis for csPCa detection, as well as a lightweight model size and complexity. RadHop is introduced for data-driven radiomics feature extraction which uses a linear and unsupervised model. A stage-2 was also proposed to reduce the false positive rate by using the proposed anomaly map derived out of stage-1 predictions, as well as combining hand-crafted and data-driven radiomics features. Evaluation comparisons show a competitive performance standing of PCa-RadHop among other works, offered in a very compact model size with low complexity. Future extensions to this work will introduce supervision in RadHop feature extraction, without compromising its transparency or small model size. 

\printcredits

\bibliographystyle{model2-names}

\bibliography{cas-refs}



\end{document}